\documentclass[aps,pre,epsfigure,twocolumn,superscriptaddress,nofootinbib]{revtex4-1}
\usepackage[colorlinks=true,linkcolor=blue,urlcolor=blue,citecolor=blue,pdfusetitle]{hyperref}
\usepackage[utf8]{inputenc}
\usepackage[english]{babel}
\usepackage{amsmath}
\usepackage[caption = false]{subfig}
\usepackage{graphicx,epstopdf}
\usepackage{blindtext}
\usepackage[table,xcdraw]{xcolor}
\usepackage{lipsum}
\usepackage{amsfonts}
\usepackage{bbm}
\usepackage{amssymb}
\usepackage{enumerate}
\usepackage{color}
\usepackage{latexsym}
\usepackage{times,txfonts}

\newcommand{\Fcal}{\mathcal{F}}

\newcommand{\Hcal}{\mathcal{H}}

\newcommand{\Qcal}{\mathcal{Q}}

\newcommand{\Ucal}{\mathcal{U}}

\newcommand{\Wcal}{\mathcal{W}}

\newcommand{\1}{\mathbbm{1}}

\begin{document}

\title{Quantum Stirling engine based on dinuclear metal complexes}

\author{Clebson Cruz}
\email{clebson.cruz@ufob.edu.br}
\affiliation{Grupo de Informa\c{c}\~{a}o Qu\^{a}ntica e F\'{i}sica Estat\'{i}stica, Centro de Ci\^{e}ncias Exatas e das Tecnologias, Universidade Federal do Oeste da Bahia - Campus Reitor Edgard Santos. Rua Bertioga, 892, Morada Nobre I, 47810-059 Barreiras, Bahia, Brasil.}

\author{Hamid-Reza Rastegar-Sedehi }
\email{h.rastegar@jahromu.ac.ir}
\affiliation{Department of Physics, College of Sciences, Jahrom University, Jahrom 74135-111, Iran}

\author{Maron F. Anka}
\email{maronanka@id.uff.br}
\affiliation{Instituto de F\'{i}sica, Universidade Federal Fluminense, Av. Gal. Milton Tavares de Souza s/n, 24210-346 Niter\'{o}i, Rio de Janeiro, Brasil.}

\author{Thiago R. de Oliveira}
\email{troliveira@id.uff.br}
\affiliation{Instituto de F\'{i}sica, Universidade Federal Fluminense, Av. Gal. Milton Tavares de Souza s/n, 24210-346 Niter\'{o}i, Rio de Janeiro, Brasil.}

\author{Mario Reis}
\email{marioreis@id.uff.br}
\affiliation{Instituto de F\'{i}sica, Universidade Federal Fluminense, Av. Gal. Milton Tavares de Souza s/n, 24210-346 Niter\'{o}i, Rio de Janeiro, Brasil.}
\affiliation{Dpto. F\'{i}sica de la Materia Condensada, Universidad de Sevilla, Apdo 1065, 41080 Sevilla, Spain}

\begin{abstract}
Low-dimensional metal complexes are versatile materials with tunable physical and chemical properties that make these systems promising platforms for caloric applications. In this context, this work proposes a quantum Stirling cycle based on a dinuclear metal complex as a working substance. The results show that the quantum cycle operational modes can be managed when considering the change in the magnetic coupling of the material and the temperature of the reservoirs. Moreover, magnetic susceptibility can be used to characterize the heat exchanges of each cycle step and, therefore, its performance. As a proof of concept, the efficiency of the heat engine is obtained from experimental susceptibility data. These results open doors for studying quantum thermodynamic cycles by using metal complexes; and further the development of emerging quantum technologies based on these advanced materials.
\end{abstract}

\maketitle

\section{Introduction}

In the past few decades, the rise of several technologies based on the quantum properties of advanced materials has attracted the attention of the scientific community \cite{wasielewski2020exploiting,gaita2019molecular,mezenov2019metal,sato2016dynamic,cruz2021quantum}. Quantum correlations and coherence have been extensively explored as primary resources in this novel quantum devices such as quantum transistors~\cite{Marchukov:16,dePonte:19} and batteries~\cite{Santos:20c,PRL_Andolina,Baris:20,Santos:19-a,cruz2021quantum}. Furthermore, the fast development of these technologies has led the literature to increasingly consider thermodynamic aspects for describing quantum processes, leading to the development of the \textit{quantum thermodynamics} area \cite{binder2018thermodynamics,vinjanampathy2016quantum}.

Analogous to the development of the classical thermodynamics, the study of heat engines have become widely explored in the  area of quantum thermodynamics \cite{Shi_2020,chatterjee2021temperature,Chand17, Hewgill17, Cakmak17, Altintas19, Johal21, Cakmak22, Maron:21, myers2022quantum, Purkait22, Solfanelli20,Pena20,Deffner18,deffner,sur2022,stirling,stirling1,stirling2,stirling3,stirling4,stirling5,papadatos2022quantum}. In particular, coupled spin systems appear as robust platforms for the development of the so-called quantum heat engines (QHEs) prototypes \cite{Cakmak17,Altintas19,Johal21,Cakmak22,Maron:21,myers2022quantum,stirling}. In this class of materials, low-dimensional molecular magnetic systems as quantum antiferromagnets \cite{mario,REIS2020100028,lloveras2021advances,CIRILLO2022101380,cruz,cruz2017influence,mario2,souza,cruz2020quantifying,he2017quantum,he2017quantum,breunig2017quantum} appear as promising platforms for quantum technologies \cite{wasielewski2020exploiting,cruz2021quantum}. In these advanced materials, intramolecular interactions are strong enough to suppress external and intermolecular interactions \cite{cruz,souza,mario2,reis2020magnetocaloric}. As a consequence, its quantum features presents highly stability against external perturbations such as high temperatures \cite{cruz,souza,mario2,cruz2021quantum}, magnetic fields \cite{cruz2020quantifying,souza2} and pressure \cite{cruz2017influence,cruz2020quantifying}. Moreover, its quantum coherence and correlations are related to substantial structural and macroscopic properties \cite{cruz,cruz2017influence,moreno2018molecular,cruz2020quantifying}, which allows the management and control of its quantum properties by experimental techniques \cite{cruz,cruz2017influence,cruz2020quantifying,sato2016dynamic}. 

Metal complexes has been explored in the literature for several caloric applications through classical thermodynamic cycles \cite{reis2020magnetocaloric,REIS2020100028,lloveras2021advances,CIRILLO2022101380}. In this scenario, this work analyzes the realization of a quantum Stirling engine whose working medium consists of a dinuclear metal complex in a d$^9$ electronic configuration. The cycle operates between two heat baths and two magnetic couplings, handled by an external hydrostatic pressure change.

The results show that the four operation modes - heat engine, refrigerator, accelerator and heater - allowed by the second law of thermodynamics can be tuned by the control of the parameters of the cycle. In addition, the heat exchanged during the cycle is analytically characterized in each cycle step through the working substance's dimensionless magnetic susceptibility. Thus, from an experimental point of view, this work also shows a proof of concept for the presented theoretical model, where the efficiency of the heat engine operational mode is obtained through magnetic susceptibility data. The presented results can provide insights for future research regarding the caloric properties of metal complexes, opening doors for studying the caloric properties of metal complexes through quantum thermodynamic cycles. These can be used in the application of dinuclear metal complexes as quantum heat engines in terms of their operation modes. Therefore, this paper lays the groundwork for future research into applying metal complexes as working substances of quantum heat engines toward the development of emerging quantum technologies based on these advanced materials. 

\section{Dinuclear metal complex as working substance}
\label{substance}

Let us consider a dinuclear metal complex in a d$^9$ electronic configuration as the working substance to a four-stroke Quantum Stirling cycle \cite{myers2022quantum, Purkait22, chatterjee2021temperature,stirling,stirling1,stirling2, stirling3,stirling4,stirling5,sur2022,papadatos2022quantum}.
This class of materials can be defined as an ideal realization of coupled 1/2-spins \cite{cruz2017influence,cruz2020quantifying,cruz2021quantum}. The Hamiltonian of this working substance is given by the Heisenberg interaction $\Hcal = J \vec{S}_1\cdot\vec{S}_2$, where $S_{n}^{(k)}\!=\!{(1/2)}\sigma_{n}^{k}$ are the spins of the d$^9$ metallic centers, with  $\sigma_{n}^{k}$ being the Pauli matrices ($k\!\in\!\{x,y,z\}$), and $J$ the magnetic coupling constant: $J>0$ corresponds to an antiparallel alignment, with entangled ground state (EGS); and $J<0$ corresponds to a parallel alignment, with separable ground state (SGS). The energy levels consist in $s=0$ singlet ($J>0$) and $s=1$ triplet ($J<0$) states \cite{mario}. Diagonalizing $\Hcal$, one can obtain the energy eigenvalues $E_s$ and the corresponding eigenstates $\vert s, m_s \rangle$ \cite{mario} as
\begin{align}
&E_{s=1}=\frac{1}{4}J  \rightarrow \left\{ 
\begin{aligned} &\vert s=1,m_s=+1\rangle =\vert \uparrow\uparrow \rangle \\
&\vert s=1,m_s=0\rangle =\frac{1}{\sqrt{2}}\left( \vert \uparrow\downarrow\rangle + \vert \downarrow\uparrow\rangle\right) \\
&\vert s=1,m_s=-1\rangle =\vert \downarrow\downarrow\rangle \end{aligned}\right. ~,\label{eq:01} \\
& E_{s=0}=-\frac{3}{4}J \rightarrow \vert s=0, m_s=0\rangle =\frac{1}{\sqrt{2}}\left( \vert \uparrow\downarrow\rangle {-} \vert \downarrow\uparrow\rangle\right) ~. 
\label{eq:02}
\end{align}

Besides the energy levels, the populations $\varrho_n$ and, consequently, the entropy have a pivotal role in quantum thermodynamic processes \cite{PhysRevE.76.031105,PhysRevE.79.041129}.
For the working substance at thermal equilibrium, the density matrix for the coupled system is given by the Gibbs form $\rho (T) = e^{-\Hcal/k_BT}/\mbox{Tr}{e^{-\Hcal/k_BT}}$, which can be written in terms of the populations $\varrho_n$, on the local basis $\lbrace \vert \uparrow \uparrow\rangle,\vert \uparrow\downarrow\rangle,\vert \downarrow\uparrow\rangle,\vert \downarrow \downarrow\rangle\rbrace$ as the X-shaped matrix:
\begin{align}
\rho &=\dfrac{1}{2}\left[
\begin{matrix} 2\varrho_2 & 0 & 0 & 0 \\
0& {\varrho_1 + \varrho_3} & {\varrho_3 - \varrho_1}  & 0\\
0& {\varrho_3 - \varrho_1}  & {\varrho_1 + \varrho_3} & 0\\
0& 0& 0 & 2\varrho_4 
\end{matrix} \right].
\label{eq:06}
\end{align}
Thus, from the system Hamiltonian $\Hcal$ and the spectral decomposition of the state $\rho$, the working substance state eigenvalues $\varrho_{n}$ (population) and its corresponding eigenvectors $\vert {\varrho_{n}}\rangle$ are written as a function of the molar magnetic susceptibility of the substance \cite{cruz,cruz2017influence,cruz2020quantifying} as
\begin{align}
&\varrho_1 = \varrho_2 = \varrho_3 = \frac{k_BT\chi(J,T)}{2N_Ag^2\mu_B^2} \rightarrow \left\{ 
\begin{aligned} &\vert {\varrho_{1}} \rangle  =\vert \uparrow\uparrow\rangle \\
&\vert {\varrho_{2}} \rangle  =\frac{1}{\sqrt{2}}\left( \vert \uparrow\downarrow\rangle + \vert \downarrow\uparrow\rangle\right) \\
&\vert {\varrho_{3}} \rangle =\vert \downarrow\downarrow\rangle \end{aligned}\right.~, \label{eq:07} \\
&\varrho_4= 1 - 3\varrho_1(T) \rightarrow  vert {\varrho_{4}} \rangle = \frac{1}{\sqrt{2}}\left( \vert \uparrow\downarrow\rangle {-} \vert \downarrow\uparrow\rangle\right) ~,
\label{eq:08}
\end{align}
where $N_A$ is the Avogadro's number, $g$ is the Land\'{e} factor, $\mu_B$ is the Bohr magneton, and $\chi$ is the Bleaney-Bowers molar magnetic susceptibility \cite{cruz,cruz2017influence,cruz2020quantifying,mario,mario2,brandao2009magnetic,bleaney1952anomalous,yurishchev2011quantum,souza} of the compound
\begin{equation}
\chi (J,T) =\frac{2 N_A(g\mu_B)^2}{k_B T}\frac{1}{3+e^{{J}/{k_B T}}}~.
\label{eq:09}
\end{equation}

From Eqs. \eqref{eq:07} and \eqref{eq:08} one can obtain the von Neumann entropy $S= - \Sigma_{i=1}^{4} \varrho_i \ln{\varrho_i} $ of the system, given by:
\begin{equation}
S(J,T) =  -\Fcal(J,T)  \left\{ e^{\frac{J}{k_BT}}\ln \left[1 - 3\Fcal(J,T)\right]  + 3 \ln \left[ \Fcal(J,T)\right]\right\}~,
\label{eq:10}
\end{equation}
where $\Fcal(J,T)$ is the so-called dimensionless magnetic susceptibility \cite{mario}, which is defined in terms of the Bleaney-Bowers equation:
\begin{equation}
    \Fcal(J,T) = \frac{k_B T\chi (J,T)}{2 N_A(g\mu_B)^2} = \frac{1}{3+e^{{J}/{k_B T}}}~.
    \label{eq:09-2}
\end{equation}

In particular, for dinuclear metal complexes, it is well known that its magnetic coupling can be handled by applying an external pressure \cite{cruz2017influence,cruz2021quantum,cruz2020quantifying,crawford1976relation,prescimone2010high,lloveras2021advances,Romanenko_2022}. {Studying the magnetic properties of metal complexes when the compound are subjected to an external hydrostatic pressure is the main purpose of the high-pressure magnetometry area \cite{prescimone2010high,PhysRevB.99.014417,fu2020sensitive}. When a high pressure is applied to a metal complex, it is possible for changes to occur in the electron density, bonding, and crystal structure of the system \cite{fu2020sensitive}.} As presented in ref.~\onlinecite{cruz2017influence}, the third-order Birch-Murnaghan equation \cite{Birch,min8120591,ma13040978} relates the magneto-structural change in metal complexes with the pressure to which the substance is subjected. Thus, a hydrostatic pressure allows one to control the magnetic coupling, which then controls the degree of quantum correlations in low dimensional metal complexes \cite{cruz2017influence}. Hence, one can manipulate the magnetic coupling ($J$) of the working substance during a thermodynamic process. Accordingly, the external pressure induces a change in the lattice parameter, and the bridging angles \cite{crawford1976relation,prescimone2010high,lloveras2021advances,Romanenko_2022}; and, as a consequence, the magnetic coupling constant can be handled, which allows the control of the energy levels. {Therefore, this approach allows the investigation of the relationship between the structure and magnetic properties of metal complexes.}
In references \cite{crawford1976relation,prescimone2010high}, the authors studied several copper dimers materials and analyzed the magnetic coupling constant as a function of the bridging angles. 
They found that a dependence of this constant with the \textit{metal-oxygen-metal} angle in metal complexes, as below:
\begin{equation}\label{Jtheta}
    J/k_{B} = 106 \theta  - 10387  \left[ K\right].
\end{equation}
Note the critical angle that makes $J=0$ is $\theta = 97.5^o$ (see references \onlinecite{crawford1976relation,prescimone2010high}).

The values of applied pressures to achieve such angles are of the order of GPa \cite{prescimone2010high,cruz2017influence}, which are experimentally achieved using commercial diamond anvil cells.
It can therefore be assumed that an antiparallel alignment of the spins is expected for  $\theta > 97.5^{o}$, while a parallel alignment is expected for  $\theta < 97.5^{o}$ \cite{prescimone2010high}. In other words, the coupling constant changes from positive (singlet state - EGS) to negative (triplet subspace - SGS) depending on the changing of the bridging angles. Thus, the magnetic coupling constant of the working substance can be {handled} during the isochoric processes by regulating the external pressure.

\section{Quantum Stirling Cycle in a metal complex}
\label{model}

A quantum thermodynamic cycle is a quantum counterpart of a classical thermodynamic cycle. While Carnot and Otto cycles have been exhaustively studied in the past few years, a wide range of other cycles can be generalized for quantum working substances \cite{myers2022quantum}. In particular, the Stirling cycle has attracted the attention of the scientific community as an alternative for the adiabatic cycles \cite{myers2022quantum,Purkait22,chatterjee2021temperature,stirling,stirling1,stirling2,stirling3,stirling4,stirling5,papadatos2022quantum}. The Stirling cycle is characterized by four main processes: (\textit{i}) isothermal expansion, (\textit{ii}) isochoric cooling, (\textit{iii}) isothermal compression, and (\textit{iv}) isochoric heating. 

There is an ongoing debate about the best definition of work for microscopic quantum systems \cite{de2020unraveling,juan2021first}. Here we will use the most common definition \cite{PhysRevE.76.031105,binder2018thermodynamics,vinjanampathy2016quantum} considering work as the (average) energy exchange due to external changes in the system's Hamiltonian. 
In this scenario, for spin systems this change can be achieved through varying an external magnetic field \cite{stirling,Purkait22}, or the magnetic coupling $J$ between the spins, as we will consider here.

Thus, the quasi-static quantum Stirling cycle can be implemented by two quantum isothermal and two quantum isochoric strokes. In this regard, the properties of the quasi-static quantum isochoric and isothermal processes are described below.

\subsection{Quantum Isochoric Processes}

The working substance is kept in contact with a thermal reservoir in the quasi-static isochoric processes. No work is performed; meanwhile, heat is exchanged between the substance and the reservoir. 
Thus, from the first law of thermodynamics for quantum processes \cite{stirling,PhysRevE.76.031105,myers2022quantum,Purkait22}, the work performed corresponds to the variation of the energy levels $E_s$ of the system, Eq. \eqref{eq:01} and \eqref{eq:02}. As a consequence, the generalized coordinate for the working substance, presented in the last section, can be defined as the magnetic coupling constant $J = E_{s=1} - E_{s=0}$, which in turn yields the change of the energy levels of the system. Therefore, considering this metal complex as the working substance, the magnetic coupling does not change during the isochoric process while the populations and the temperature change.

\subsection{Quantum Isothermal Processes}

The isothermal processes of the quantum Stirling cycle can be summarized as the working substance in equilibrium with the heat bath with fixed temperature $T$; while, at the same time, the generalized coordinate, given by the magnetic coupling constant of the system, is changed quasi-statically, leading to a change of the system's energy spectrum $E_n$ and the corresponding populations $\varrho_{n}$ \cite{PhysRevE.76.031105}. 

Work can be performed or received in this process, while the working substance exchanges heat with the reservoir. The magnetic coupling $J$ and populations $\varrho_n$ (Eqs. \eqref{eq:07} - \eqref{eq:08}) of the material is changed slowly due to the application of an external hydrostatic pressure (see equation \ref{Jtheta}), ensuring that the substance is kept in thermal equilibrium with the reservoir during the entire process.  

In the classical analogs, the internal energy $\Ucal$ of the system remains constant while it experiences an isothermal process. In contrast with its classical counterpart, the isothermal process for the quantum Stirling cycle has a non-zero internal energy change \cite{chatterjee2021temperature,PhysRevE.76.031105,PhysRevE.79.041129}. The internal energy of the working substance is obtained from the  $\Ucal = \mbox{Tr}\left[ \rho \Hcal\right] $, which can also be calculated as a function of the dimensionless magnetic susceptibility as:
\begin{equation}
U(J,T) = 3J\left[\Fcal(J,T) - \frac{1}{4}\right].
\label{eq:15}
\end{equation}
In this context, for the dinuclear metal complex working substance, the internal energy of the system is directly related to its magnetic coupling constant and, consequently, sensitive to external pressure\cite{cruz2017influence}.

\subsection{Step-by-step of the quantum Stirling cycle}

\begin{figure*}
    \centering
    \includegraphics[width=15cm]{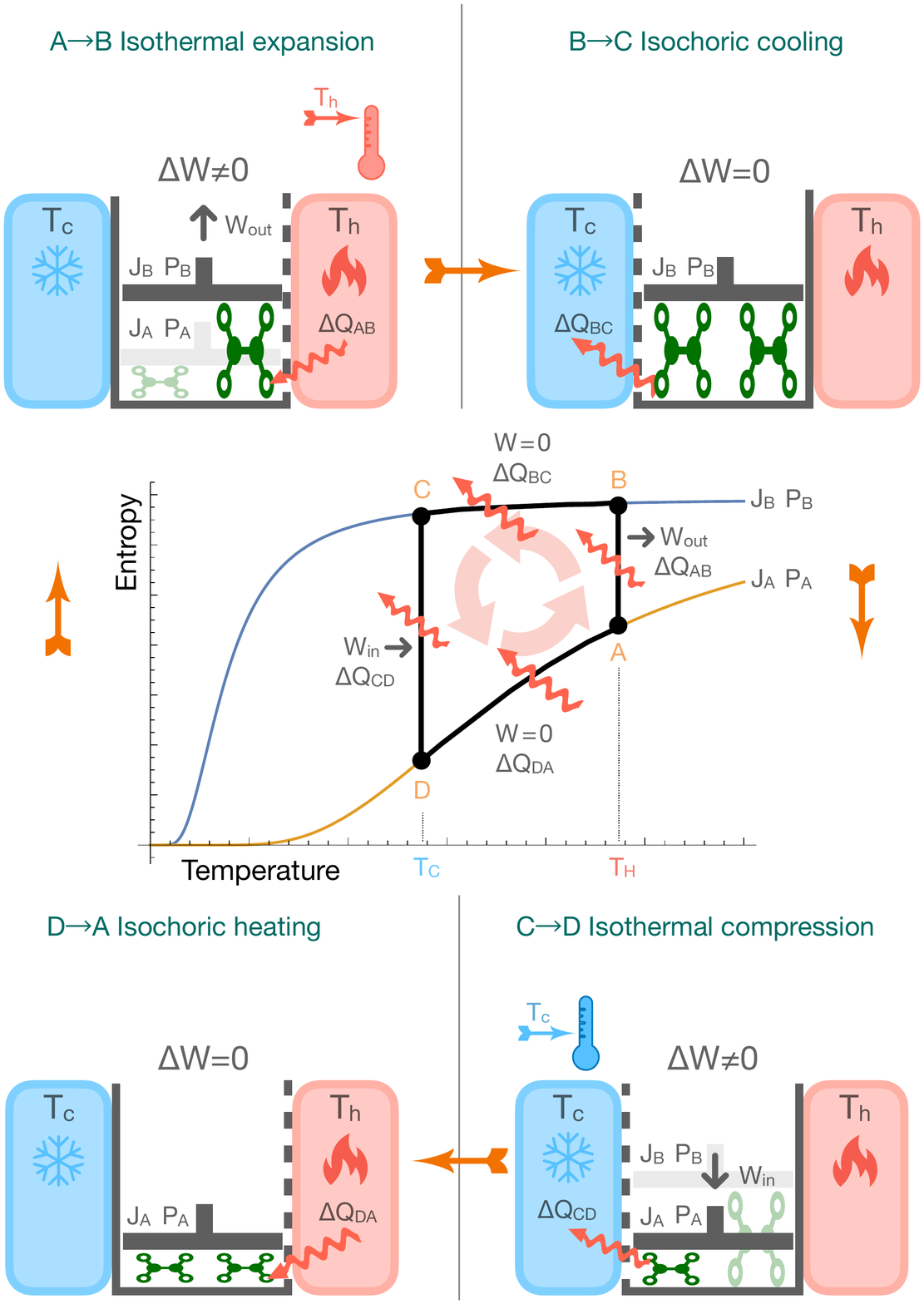}
    \caption{Sketch of the four-stage quantum Stirling cycle operating as heat engine, using a dinuclear metal complex as a working substance. The engine operates between two magnetic couplings $J_{A}$ and $J_{B}0$, induced by the applying of an external pressure, and two thermal baths $T_{h}$ (red) and $T_{c}$ (blue), with  $T_{hot} > T_{cool}$. The quantum cycle is composed of two isothermal ($A\rightarrow B$ and $C\rightarrow D$) and two isochoric ($B\rightarrow C$ and $D\rightarrow A$) strokes. The system will absorb ($Q >0$) or release ($Q < 0$) heat regarding the signal of these magnetic coupling. By convention, $W<0$ and $W>0$ corresponds to the work done \textit{on} and \textit{by} the system respectively.}
    \label{fig:02}
\end{figure*}

The principal subject of study in any cyclic heat engine is the characterization of the heat exchanged and the work performed in one complete cycle. 
Based on the working substance and the thermodynamic processes presented before, one can implement the four-stage quantum Stirling cycle and study its properties. Fig. \ref{fig:02} shows a sketch of the steps and the $ST$  diagram for this cycle. As it can be seen, the engine operates between two heat baths, with temperatures $T_{h}$ and $T_{c}$ (with $T_{h}>T_{c}$); and two different pressure values, meaning two different magnetic couplings  $J_{A}$ and $J_{B}$. 

{It is important to highlight that, in order to ensure that the system is in the low-temperature regime where the ground state is primarily occupied, and the quantum effects are significant, the cycle needs to be implemented far enough from the Curie's paramagnetic region ($k_BT\gg \vert J\vert$) \cite{cruz,mario}. Therefore, the results presented in this section were obtained under the condition that the temperature of the thermal baths are distant from the Curie regime and the cycle operates at low-temperature.}


\subsubsection{Isothermal expansion ($A\rightarrow B$)}

In the first step of the cycle, the working substance is in thermal equilibrium with a reservoir at temperature $T_{h}$, while it experiences a quasi-static expansion, i.e., pressure is then released from $P_A$ to $P_B$, changing its magnetic coupling from $J_{A}$ to  $J_{B}$. This change slowly modifies the energy levels of the system in equilibrium with the thermal bath \cite{PhysRevE.76.031105,PhysRevE.79.041129}. 
The change in internal energy is not zero, which is a pivotal difference from the classical cycle, where the internal energy stays constant during an isothermal process. The heat exchanged with the reservoir at temperature $T_h$, during this process can be found in terms of the von Neumann entropy change \footnote{In the quantum isothermal, as in the quantum adiabatic, the system is usually driven out of thermal equilibrium. It is only for systems where all the energy levels are equidistant that the equilibrium is preserved. As we have degeneracy and thus only two levels the
equilibrium is preserved in the isothermal and we can obtain the heat from entropy variation.}:
\begin{equation}
\Delta Q_{AB}= \int_{A}^{B} T_{h} dS = T_{h} \left[ S(J_{B},T_h) - S(J_{A},T_h)\right],
\label{qab}
\end{equation}
where the entropy $S(J_i,T_j)$ is given by Eq. \eqref{eq:10}.

\begin{figure}
    \centering
{\includegraphics[width=8.5cm]{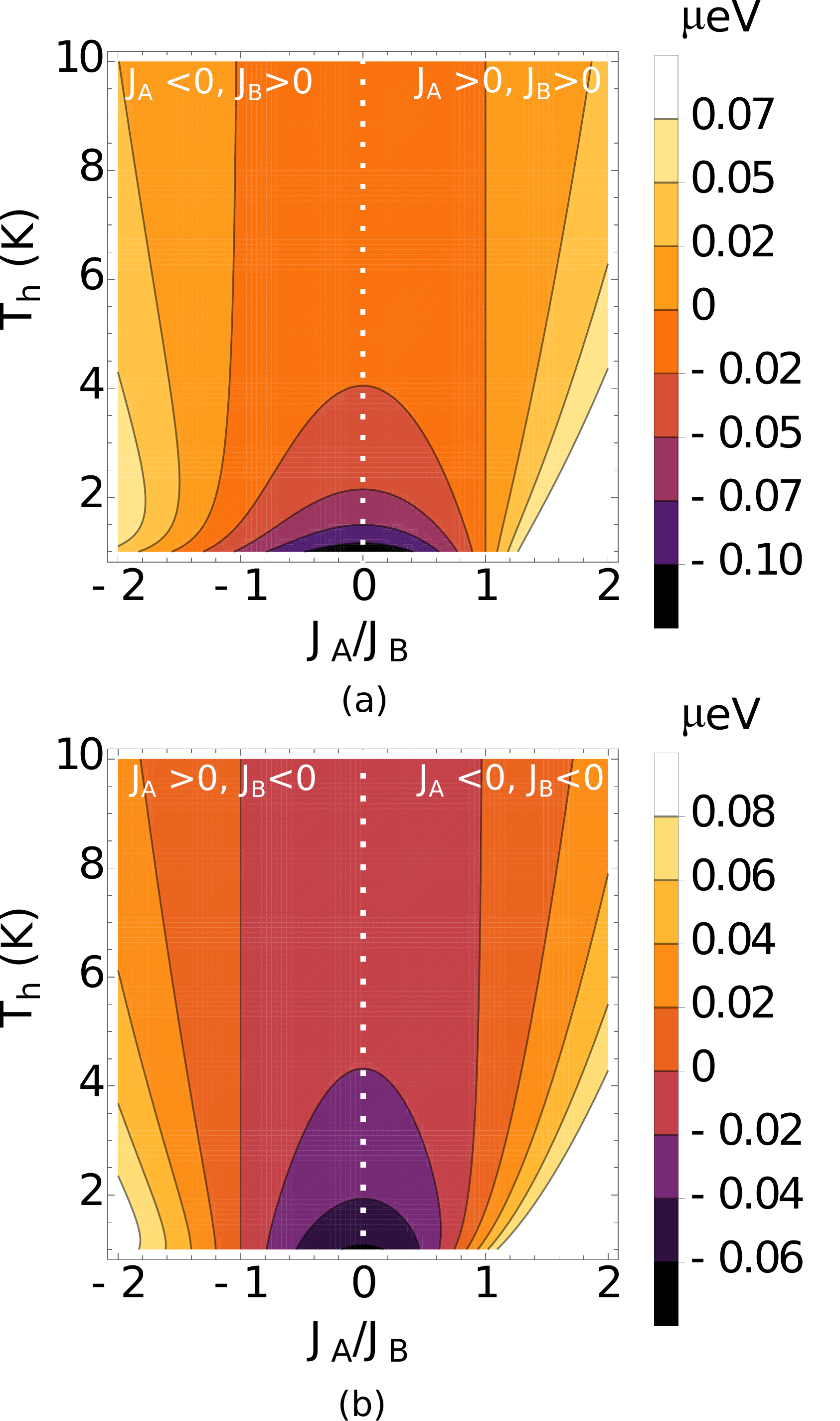}}
    \caption{(Color online) Heat exchanged with the hot reservoir during the isothermal expansion. Dotted white line separates the different physical scenarios regarding to the magnetic coupling: (a) $\{J_{A} > 0, {J_{B} >0\}}$ - right and $\{ J_{A} < 0, J_{B}>0\}$ - left; (b) $\{J_{A} < 0, J_{B}<0\}$ - right and $\{J_{A} > 0,J_{B}<0\}$ - left. Figure (a) represents a working substance that increases the magnetic coupling during an expansion while (b) reduces it. {The selected coupling ratio scale is based on the experimental feasible region, which can be achieved by applying an external hydrostatic pressure to change the magnetic coupling of s-1/2 metal complexes \cite{prescimone2010high}.}}
    \label{fig:qab}
\end{figure}

Figure \ref{fig:qab} shows the heat exchanged during the isothermal expansion. As it can be seen, the working substance can absorb ($\Delta {Q}>0$) or release ($\Delta {Q}<0$)  heat from the hot bath depending on the ratio between the magnetic coupling of the working substance before ($J_{A}$) and after ($J_{B}$) the expansion. Since the Stirling engine absorbs heat in this stroke, this change indicates that the quantum cycle can operate in different regions as described in Section \ref{operation}.  {Thus, expressing this result in terms of the ratio between the magnetic couplings allows the tuning of the operational modes of the engine.}

\subsubsection{Isochoric cooling ($B\rightarrow C$)}

In this step, the system is kept at constant pressure $P_B$ in order to keep the magnetic coupling constant $J_{B}$ unchanged, as well as its energy levels. Meanwhile, the system is quasi-statically connected to the bath at temperature $T_{c}<T_{h}$. Thus, as its classical counterpart in the isochoric cooling, no work is done, and the system loses heat. Therefore, the heat lost in this process calculated from the first law of thermodynamics \cite{PhysRevE.76.031105} is equivalent to the change in the internal energy of the thermal state: 
\begin{eqnarray}
    \Delta Q_{BC} &=& U(J_{B},T_{c}) -U(J_{B},T_{h})\nonumber\\
&=& 3 J_{B} [\Fcal(J_{B},T_{c}) - \Fcal(J_{B},T_{h}) ]
    \label{qbc}
\end{eqnarray}

\begin{figure}[h!]
    \centering
\includegraphics[width=8cm]{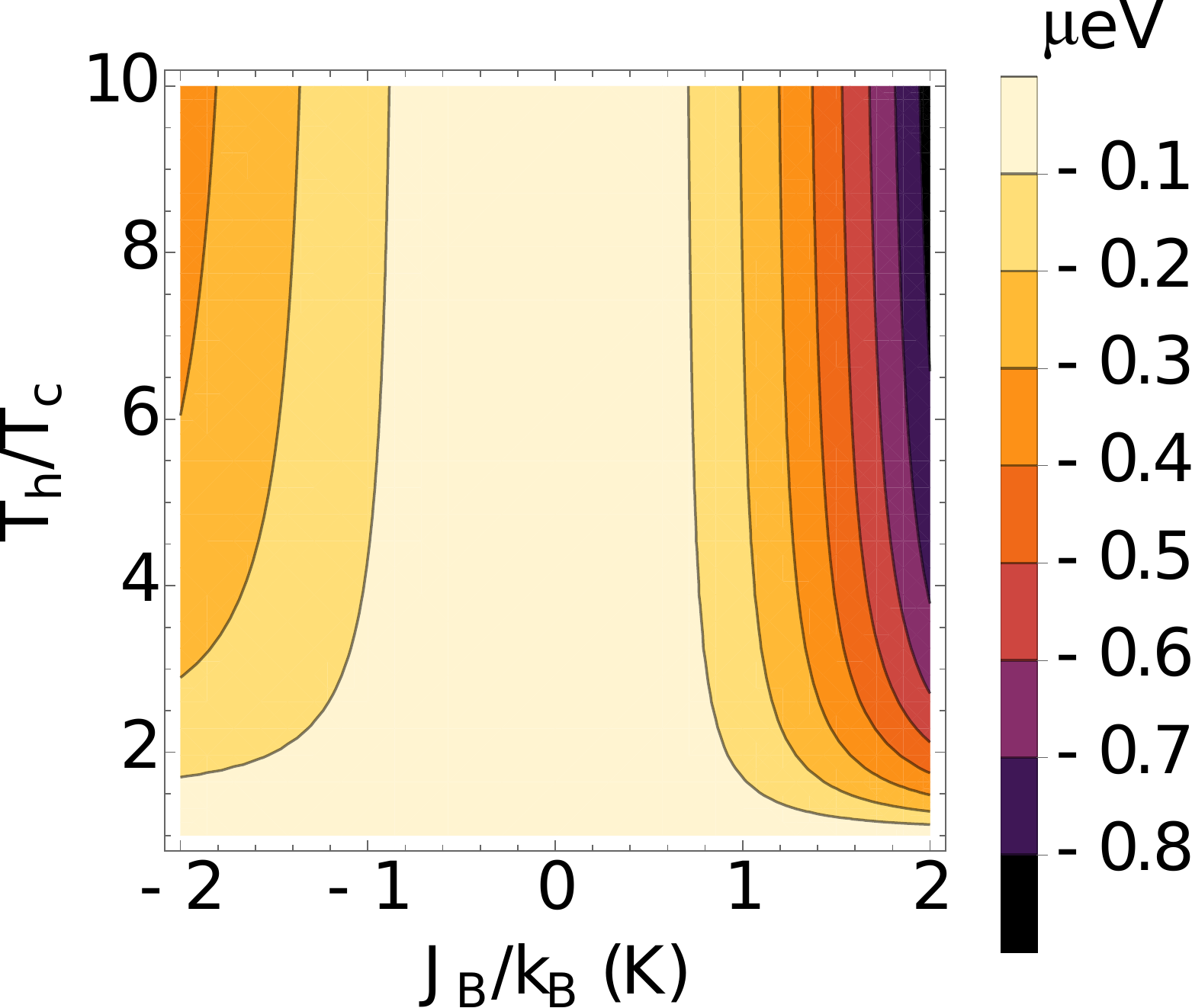}
    \caption{(Color online) Heat exchanged during the isochoric cooling. As can be seen, the working substance only releases heat ($\Delta {Q}<0$) during this stroke.}
    \label{fig:qbc}
\end{figure}

Figure \ref{fig:qbc} shows the plot of Eq. \eqref{qbc} in terms of the ratio between the temperatures of the reservoirs ($T_{h}/T_{c}$), and the magnetic coupling ($J_{B}$) in the isochoric cooling. {In contrast to the isothermal strokes, the temperatures of the hot and cold baths do not necessarily need to be fixed to obtain Figure \ref{fig:qbc}. It is only necessary to ensure that their ratio is appropriately set. However, it is still important to ensure that the system is far enough from Curie’s paramagnetic region ($k_BT\gg \vert J\vert$) \cite{cruz}.} It is worth noting that, as expected from the classical cycle, the system releases heat ($\Delta {Q}<0$) during the isochoric cooling regardless of the parameters of the cycle during this stroke.

\subsubsection{Isothermal compression ($C\rightarrow D$)}

Following the cycle, the material is then placed in thermal equilibrium with the reservoir at temperature $T_{c}$. The pressure 
is then applied from $P_B$ to $P_A$, and the magnetic coupling changes from $J_B$ to $J_A$.
Therefore, similarly to step $A\rightarrow B$, 
the heat exchanged in this step is found in terms of the von Neumann entropy change: 
\begin{equation}
\Delta{Q}_{CD}= \int_{C}^{D} T_{c} dS = T_{c} \left[ S(J_{A},T_c) - S(J_{B},T_c)\right].\label{qcd}
\end{equation}

\begin{figure}[h!]
    \centering
\includegraphics[width=8.5cm]{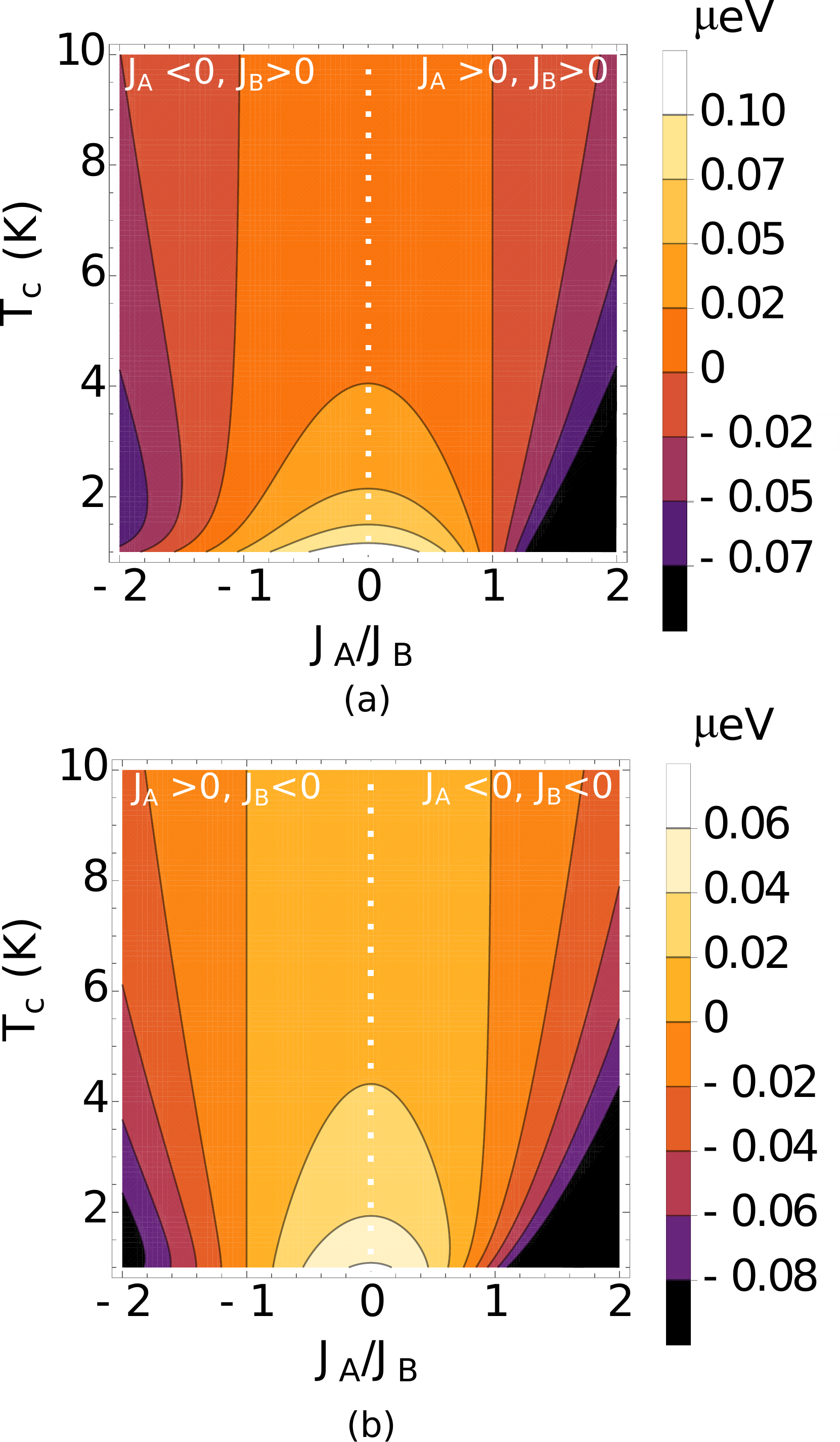}
    \caption{(Color online)  Heat exchanged with the cold reservoir during the isothermal compression. In a similar fashion as done in Fig. \ref{fig:qab}, dotted white line separates the different physical scenarios regarding to the magnetic coupling: (a) $\{J_{A} > 0, J_{B}\}>0$ - right and $\{ J_{A} < 0, J_{B}>0\}$ - left; (b) $\{J_{A} < 0, J_{B}<0\}$ - right and $\{J_{A} > 0,J_{B}<0\}$ - left. Figure (a) corresponds to a working substance that increases the magnetic coupling during an expansion while (b) reduces it.}
    \label{fig:qcd}
\end{figure}

Figure \ref{fig:qcd} shows the heat exchanged during the isothermal compression. It is observed that the system can absorb or releases heat in the cold bath depending on the parameters of this stroke. {In addition, as done in Fig. \ref{fig:qab}, by expressing the result in terms of the ratio between the magnetic couplings ($J_{A}/J_{B}$), the values of $J_{A}$ or $J_{B}$ do not need to be fixed. However, the temperature of the cold bath needs to be defined in order to ensure that the stroke is performed far enough of the Curie’s paramagnetic region ($k_BT\ll \vert J\vert$) \cite{cruz}.}



\subsubsection{Isochoric heating ($D\rightarrow A$)}

In order to close the cycle, the working substance is removed from the contact with the bath at temperature $T_{c}$, undergoing another isochoric process. It happens under pressure $P_A$ and  magnetic coupling constant $J_{A}$. At the end of the process, the material reaches an equilibrium temperature of $T_{h}>T_{c}$ (point A). In the classical Stirling heat engine, this is the isochoric heating, where the volume remains constant, and thus no mechanical work is done - and the same happens for this quantum Stirling cycle: no work is done along this isochoric process. Thus, the heat absorbed in this process is equivalent to the internal energy change:
\begin{eqnarray}
    \Delta Q_{DA} &=& U(J_{A},T_{h}) -  U(J_{A},T_{c})\nonumber\\
&=& 3 J_{A} [\Fcal(J_{A},T_{h}) - \Fcal(J_{A},T_{c}) ]
    \label{qda}
\end{eqnarray}

\begin{figure}[h!]
    \centering
\includegraphics[width=8cm]{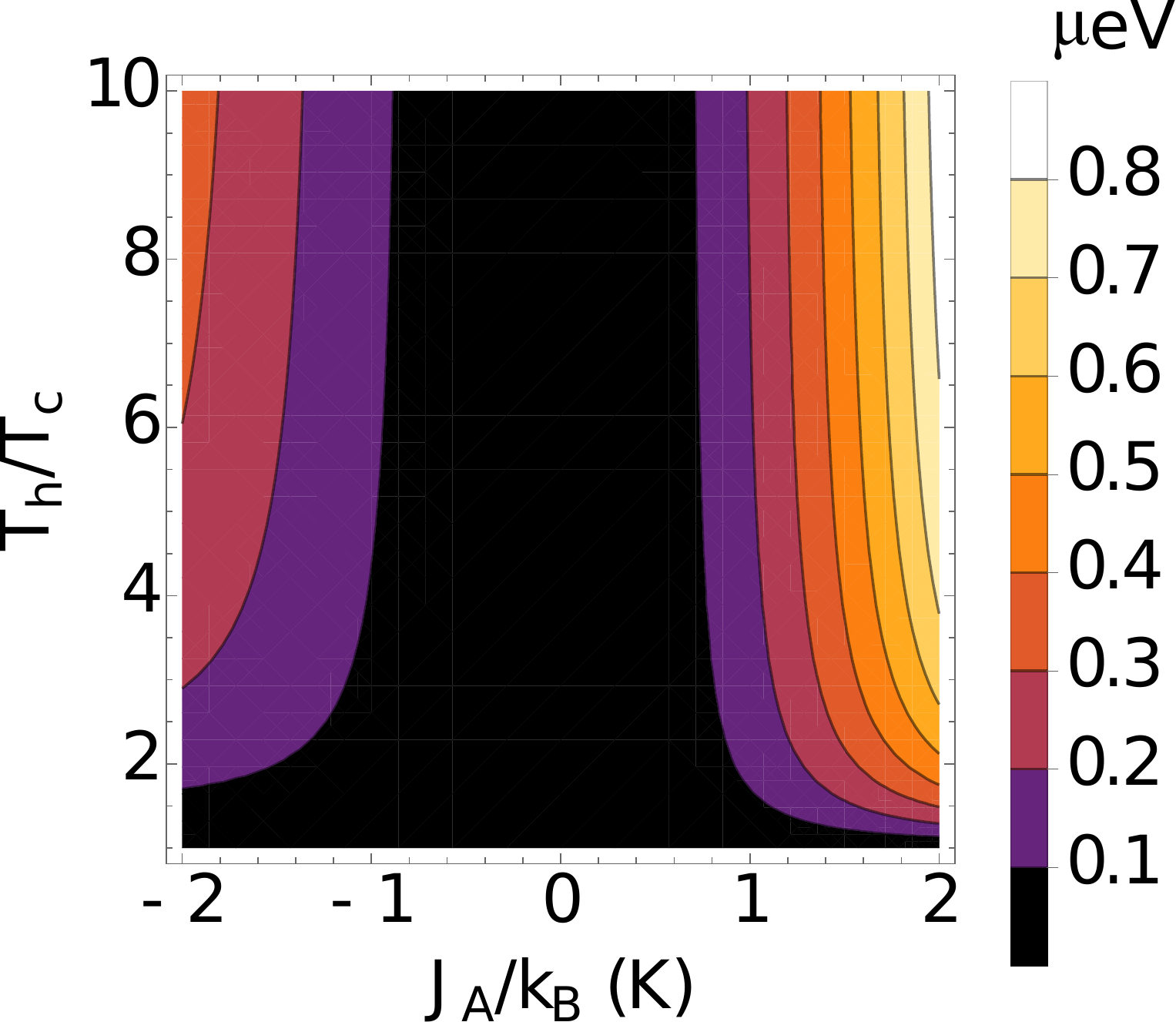}
    \caption{(Color online)  Heat exchanged during the isochoric heating. As can be seen, the working substance only absorbs heat {($\Delta {Q}>0$)} during this stroke.}
    \label{fig:qda}
\end{figure}

Figure \ref{fig:qda} shows heat exchanged during the isochoric heating. The system absorbs heat ($\Delta {Q}>0$), regardless of the magnetic coupling of the working substance during the stroke or the ratio between the temperatures of the thermal baths.

\section{Operational modes: an analysis}
\label{operation}

For this section, the operational modes are described in detail, where heat engine, refrigerator, accelerator, and heater regimes are observed, depending on the values of the magnetic coupling constants or temperature of the reservoirs. From the previous section, it is clear that similar to the classical cycle, the working substance during the $B\rightarrow C$ process (isochoric cooling) releases heat, while this material absorbs heat along the $D\rightarrow A$ process (isochoric heating). On the other hand, for the isothermal expansion ($A\rightarrow B$ step) and compression ($C\rightarrow D$ step), the working substance either absorbs or releases heat, depending on the values of magnetic coupling constants $J_A$ and $J_B$. 

Considering the classical analogs with the Stirling Cycle \cite{stirling,PhysRevE.76.031105,myers2022quantum,Purkait22}, the working substance absorbs heat as it experiences the isothermal expansion ($A\rightarrow B$), and the isochoric heating ($D\rightarrow A$). Thus, the heat absorbed during one quantum cycle is defined by:
\begin{eqnarray}
\Qcal_{in} &=& \Delta{Q}_{AB} + \Delta{Q}_{DA}.
\label{qin}
\end{eqnarray}
In the same way, the heat released in the isochoric cooling ($B\rightarrow C$) and isothermal compression ($C\rightarrow D$) is given by:
\begin{eqnarray}
\Qcal_{out} &=& \Delta{Q}_{BC} + \Delta{Q}_{CD}~.
\label{qout}
\end{eqnarray}

Considering the four thermodynamic processes described in the previous section, one can obtain the total work from the first law of thermodynamics as \cite{sur2022,chatterjee2021temperature,Purkait22}:
\begin{eqnarray}
\Wcal &=& T_{h}\left\{\ln{\left[e^{\frac{J_{A}-J_{B}}{4k_{B}T_{h}}}\frac{\Fcal(J_A,T_h)}{\Fcal(J_B,T_h)}\right]} + \frac{T_{c}}{T_{h}} \ln{\left[e^{\frac{J_{B}-J_{A}}{4k_{B}T_{c}}}\frac{\Fcal(J_B,T_c)}{\Fcal(J_A,T_c)}\right]}\right\}. \nonumber\\
\label{work}
\end{eqnarray}
The first term on Eq. \eqref{work} corresponds to the work (${W}_{AB}$) done  \textit{by} the material during the isothermal expansion ($A\rightarrow B$) and, therefore, it is expected to be positive \cite{Solfanelli20}. On the other hand, the second term corresponds to the work (${W}_{CD}$) done \textit{on} the material during the isothermal compression ($C\rightarrow D$) and hence is expected to be negative. 

\begin{figure}[h!]
    \centering
    \includegraphics[width=8cm]{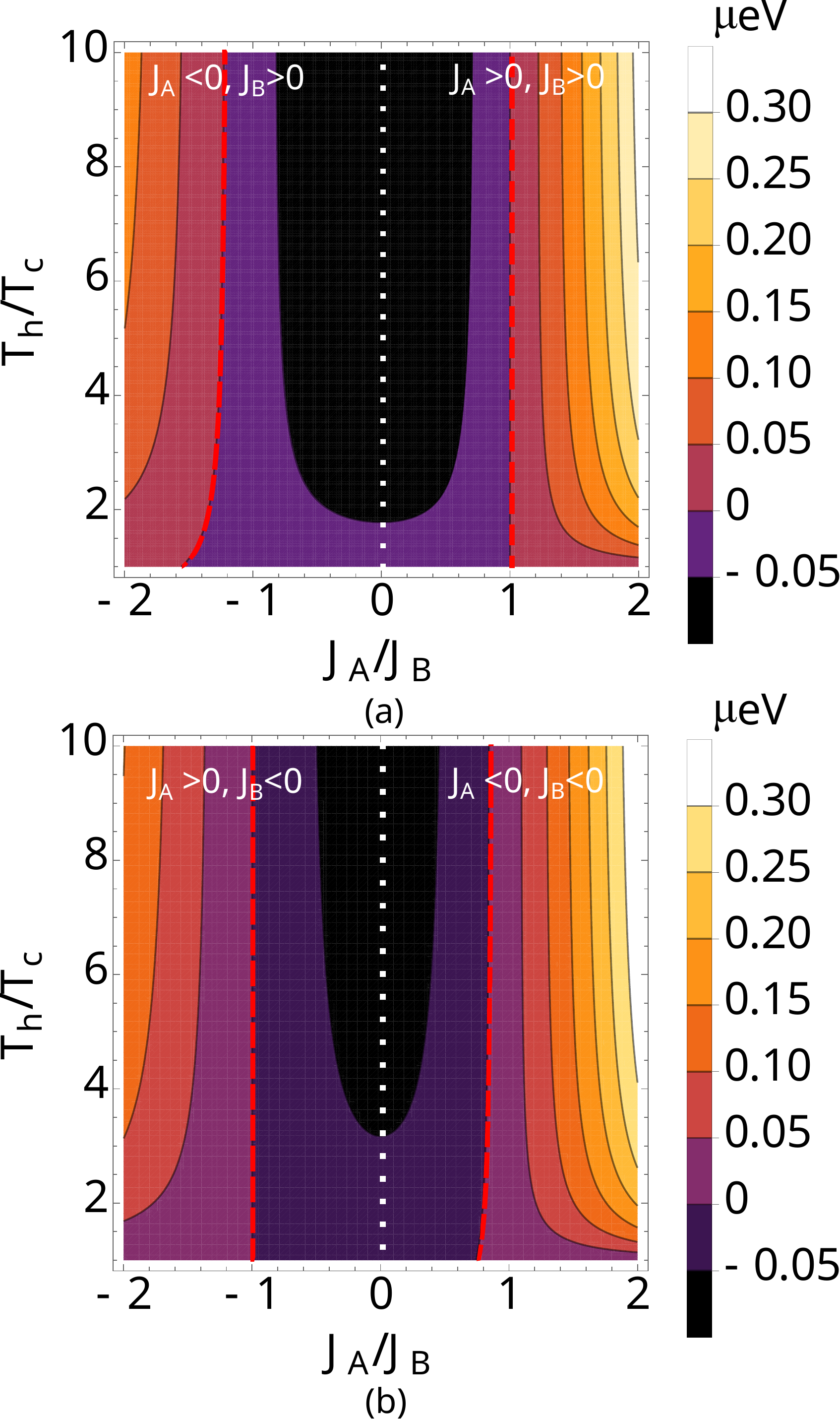}
    \caption{{(Color online) Total work for a complete quantum Stirling cycle in terms of the ratio of the parameters of the engine: temperatures ($T_{h}/T_{c}$)  and the magnetic couplings ($J_{A}/J_{B}$). The dashed red line represents the $\Wcal = 0$ border of the positive work condition. A white dashed line divides the areas for various combinations of magnetic coupling signs; these two pictures depict four distinct physical situations: (a) $\{J_{A} > 0, J_{B}>0 \}$ - right and $\{ J_{A} < 0, J_{B}>0\}$ - left; (b) $\{J_{A} < 0, J_{B}<0\}$ - right and $\{J_{A} > 0, J_{B}<0\}$ - left. In a manner comparable to that of Figs. \ref{fig:qab} and \ref{fig:qcd}, figure (a) illustrates a working material that increases magnetic coupling during an expansion, while (b) decreases it. Therefore, the discrepancy between these two figures is because the same working material cannot simultaneously increase and reduce coupling owing to expansion/compression.}}
    \label{fig:work_output}
\end{figure}

{Since the cycle operates in a quasi-static regime (no time involved), the power is zero. However, from Eq. \eqref{work}, one can estimate the total work done \textit{by/on} the system. Fig. \ref{fig:work_output} shows the total work for a complete quantum Stirling cycle in terms of the ratio of the heat baths temperatures ($T_{h}/T_{c}$)  and the magnetic couplings ($J_{A}/J_{B}$), obtained from the analytical forms of Eqs. \eqref{qin}, \eqref{qout} and \eqref{work}.}

{The dashed red line in Fig. \ref{fig:work_output} indicates the zero work curves ($\Wcal = 0$), which define the boundaries of the region where positive work is being done. As can be seen, each combination of the coupling signs has a unique zero work curve. From Figures \ref{qab} and \ref{qcd}, one can identify that the zero work curves in Fig. \ref{fig:work_output} correlate to the zero contour line in Figures \ref{qab} and \ref{qcd}, as anticipated by the first law of thermodynamics. Therefore, the amount of heat that is transferred between the isochoric strokes is equal to $\Delta{Q}_{BC} +\Delta{Q}_{DA} =0$ along the  curves, i.e., the heat that is given out during the isochoric cooling ($B\rightarrow C$) is balanced out by the heat that is taken in during the isochoric heating ($D\rightarrow A$).} 

In this regard, {by controlling the ratio between the parameters of the cycle, one can change the sign of the total work of the cycle. Thus,} from Eqs. \eqref{work}, \eqref{qin}, and \eqref{qout}, one can illustrate the regions of operation for this quantum Stirling cycle in accordance with the  second law of thermodynamics \cite{Solfanelli20,deffner}.
For the case $T_{h} > T_{c}$ the regions of operation of any thermodynamic cycle can be defined from the Clausius formulation of the second law of thermodynamics \cite{Solfanelli20,deffner,sur2022}, as in Table \ref{Clau}.
\begin{table}[h!]\caption{Operational modes for a quantum cycle allowed by the  second law of thermodynamics \cite{Solfanelli20,deffner}. The signal '+' means work done \textit{by} the system and heat absorbed; '-' means work done \textit{on} the system and heat released  \cite{Solfanelli20}.}
\begin{tabular}{l|c|c|c}
             & $\mathcal{W}$ & $\mathcal{Q}_{in}$ & $\mathcal{Q}_{out}$ \\\hline\hline
Heat engine  & + & + & -    \\\hline
Refrigerator & - & - & +    \\\hline
Accelerator  & - & + & -    \\\hline
Heater       & - & - & -   
\end{tabular}\label{Clau}
\end{table}

These modes can be described as follow:
\begin{description}
\item[Heat engine] absorbs heat from the hot bath at $T_h$ ($\Qcal_{in}>0$) and releases it into the cold one at $T_c$ ($\Qcal_{out}>0$), converting part the heat flux between these two reservoirs in work done \textit{by} the system ($\Wcal > 0$). The external work performed \textit{on} the system in the isothermal compression is smaller than the work performed \textit{by} the system in the isothermal expansion: ${W}_{AB}>{W}_{CD}$. 
\item[Refrigerator] forces a heat flow from the cold bath at $T_c$ to the hot one at $T_h$; i.e, the working substance absorbs heat from the bath at $T_c$ ($\Qcal_{out}>0$) and releases it in the one $T_h$($\Qcal_{in}<0$). For this mode, the work \textit{on} the system is greater than the work performed \textit{by} the system: ${W}_{AB}<{W}_{CD}$ ($\Wcal <  0$), which induces a heat flow from the cold bath to the hot one.
\item[Accelerator] uses the work done \textit{on} the system ($\Wcal <  0$) to enhances the heat flow from the bath at $T_h$ to the one at $T_c$ ($\Qcal_{in}>0$ and $\Qcal_{out}<0$).
\item[Heater] uses the work performed on the system ($\Wcal <  0$) to induce heat flow in both baths, i.e., the working substance releases heat in both baths ($\Qcal_{in}<0$ and ($\Qcal_{out}<0$).
\end{description}

\begin{figure}[h!]
    \centering
    \includegraphics[width=8cm]{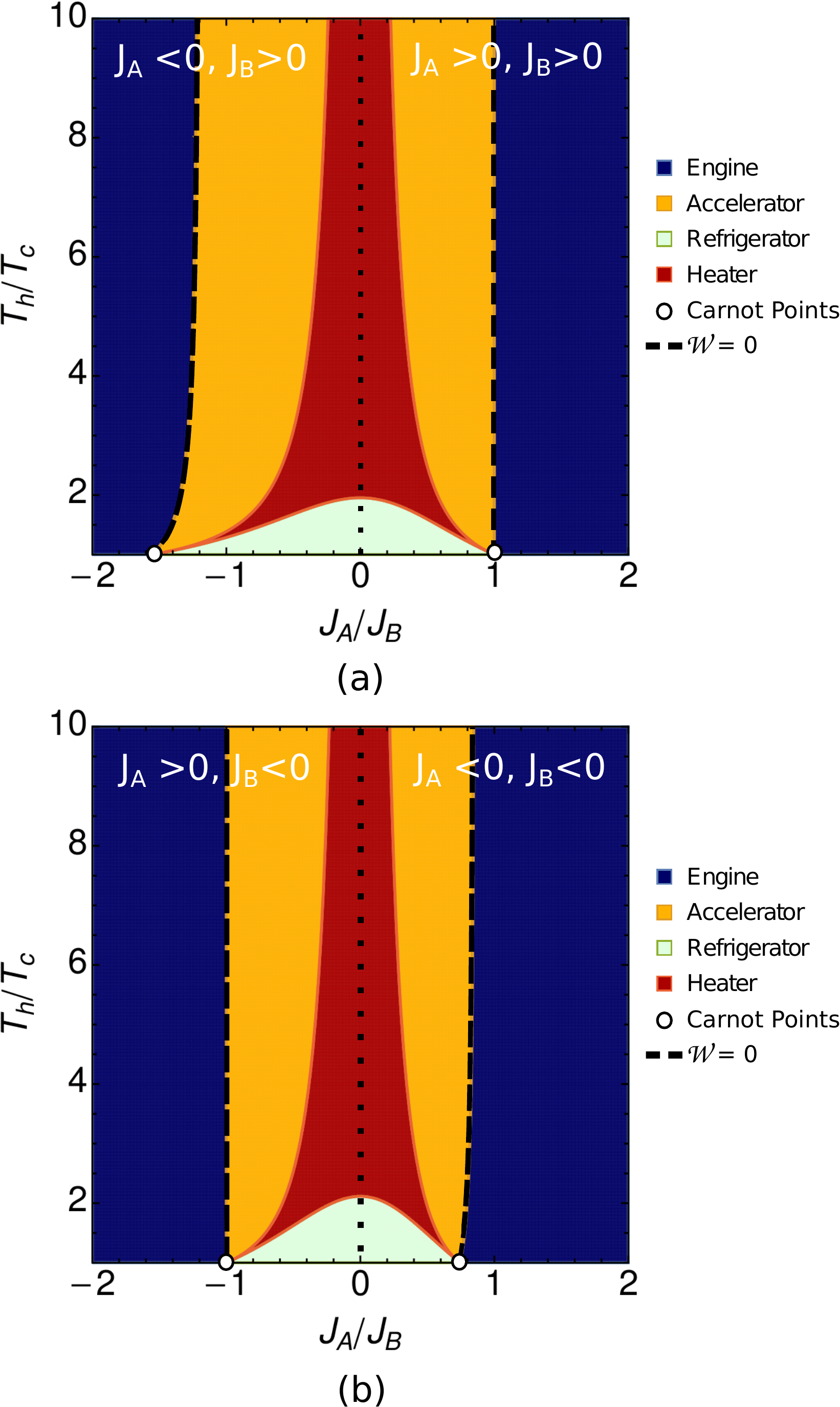}
    \caption{(Color online) Operation modes for a complete quantum Stirling cycle in terms of the ratio of the parameters of the engine: temperatures ($T_{h}/T_{c}$)  and the magnetic couplings ($J_{A}/J_{B}$). The regions corresponding to each operation mode are marked as heat engine - blue; refrigerator - green; accelerator - yellow; and heater - red. The black dashed line is the boundary $\Wcal = 0$ of the positive work condition. The open circles represent the Carnot Points, in which the four operation modes coincide. Black dotted line separates the regions for different combinations of magnetic coupling {signs; there are} four possible physical scenarios, separated in these two figures: (a) $\{J_{A} > 0, J_{B}\}>0$ - right and $\{ J_{A} < 0, J_{B}>0\}$ - left; (b) $\{J_{A} < 0, J_{B}<0\}$ - right and $\{J_{A} > 0, J_{B}<0\}$ - left. In a similar fashion as done in Figs. \ref{fig:qab} and \ref{fig:qcd}, figure (a) represents a working substance that increases the magnetic coupling during an expansion while (b) reduces it. Therefore, the difference between these two figures is due to the fact that the same working substance cannot increase and decrease coupling simultaneously due to expansion/compression.}
    \label{fig:regions}
\end{figure}

Fig. \ref{fig:regions} {illustrates the operating regions for a full quantum Stirling cycle in terms of the ratio of the heat baths temperatures ($T_{h}/T_{c}$) and the magnetic couplings ($J_{A}/J_{B}$), calculated from the analytical versions of Eqs. \eqref{qin}, \eqref{qout} and \eqref{work}.}

The four regions described above are represented in the $J_{A}/J_{B},T_{h}/T_{c}$ plane, different from the results observed for the classical analogs and the magnetic version of the quantum cycle \cite{Purkait22}, where only the heat engine and refrigerator modes are found.  Boundary lines separate the different operation regions; and crossing these boundaries means changing the sign of the respective energy exchange according to Table \ref{Clau}. The most striking result which emerges from Fig. \ref{fig:regions} is that the operation of a quantum Stirling cycle, based on a spin-1/2 dinuclear metal complex, can be tuned by the management of the ratio between different magnetic couplings ($J_{A}/J_{B}$) and the temperature of the thermal baths ($T_{h}/T_{c}$). In particular, the ratio $J_{A}/J_{B}$ leads to the control of the operational regime of the cycle through the management of the applied pressure during the isothermal compression. 

In addition, any operation mode can be achieved regardless of the presence of an entangled ground state. However, the boundaries between the operation regions can be tuned by the parameters of the cycle. 
In this regard, turning to the positive work condition, the work done on the system to change the magnetic coupling from $J_{A}$ to $J_{B}$ in the isothermal expansion is greater than the work performed in the isothermal compression. 
This region becomes predominant as we increase the ratio between the couplings regardless of the ground state. In other words, for systems in which the coupling is reduced by applying an external pressure ($J_{A}>J_{B}$) \cite{prescimone2010high}, the predominant regime is the heat engine. 

On the other hand, by analyzing the zero work curves, one can be seen that as the ratio between the temperatures of the cycle decreases, the curve reduces to a single point at $T_{h}/T_{c}=1$ where the four operation modes coincide. These points are also known as Carnot points, where all energy exchanged goes to zero (nothing happens) while the efficiency in the Heat Engine operation approaches the Carnot efficiency \cite{Solfanelli20}.
\begin{equation}
    \eta_{C} = 1 - \frac{T_{c}}{T_{h}}~.
    \label{carnot}
\end{equation}

Alternatively, in a system that does not meet positive working conditions, efficiency is no longer an important performance indicator \cite{deffner}. Apart from the heat engine mode, the regions of accelerator, refrigerator and, heater ($\Wcal < 0$) are observed for lower values of the coupling ratio in Fig. \ref{fig:regions}. In particular, the refrigerator mode is only observed by reducing the ratio between the temperature of the thermal baths. This result is consistent with the fact that fixing the temperature of the hot reservoir ($T_{h}$), extracting heat from the cold bath becomes harder as the temperature $T_{c}$ increases. In this scenario, the work done in the isothermal compression will be greater than the work performed in the isothermal expansion. 

In contrast, the fraction of the work done can induce a free flow of heat from the hot to the cold bath, which makes the system  operate as an accelerator; or reject heat in both baths as a heater. Moreover, increasing the ratio $T_{h}/T_{c}$ in the negative work region lowers the heater operational mode. These results are consistent with the intuitive idea that fixing the temperature of the cold bath ($T_{c}$), damping heat in both reservoirs becomes difficult as we increase the temperature of the hot bath ($T_{h}$), favoring the appearance of the accelerator region. 

\section{Efficiency of the heat engine mode}
The thermodynamic efficiency $\eta_{E}$ for the heat engine mode is defined as the ratio between the work extracted, Eq. \eqref{work}, and the heat absorbed, Eq. \eqref{qin}, by the working substance:
\begin{eqnarray}
    \eta &=& \frac{\Wcal}{\Qcal_{in}}\nonumber \\
    &=& \frac{\left\{ \ln{\left[e^{\frac{J_{A}-J_{B}}{4k_{B}T_{h}}}\frac{\Fcal(J_A,T_h)}{\Fcal(J_B,T_h)}\right]} + \frac{T_{c}}{T_{h}} \ln{\left[e^{\frac{J_{B}-J_{A}}{4k_{B}T_{c}}}\frac{\Fcal(J_B,T_c)}{\Fcal(J_A,T_c)}\right]}\right\}}{\left\{\left[ {S(J_B,T_h)-S(J_A,T_h)}\right]+  3 {\frac{J_{A}}{T_{h}}} [\Fcal(J_{A},T_{h}) - \Fcal(J_{A},T_{c}) ]\right\}}\nonumber\\
    \label{efficiency}
    \end{eqnarray}

\begin{figure}[h!]
    \centering
    \includegraphics[width=8cm]{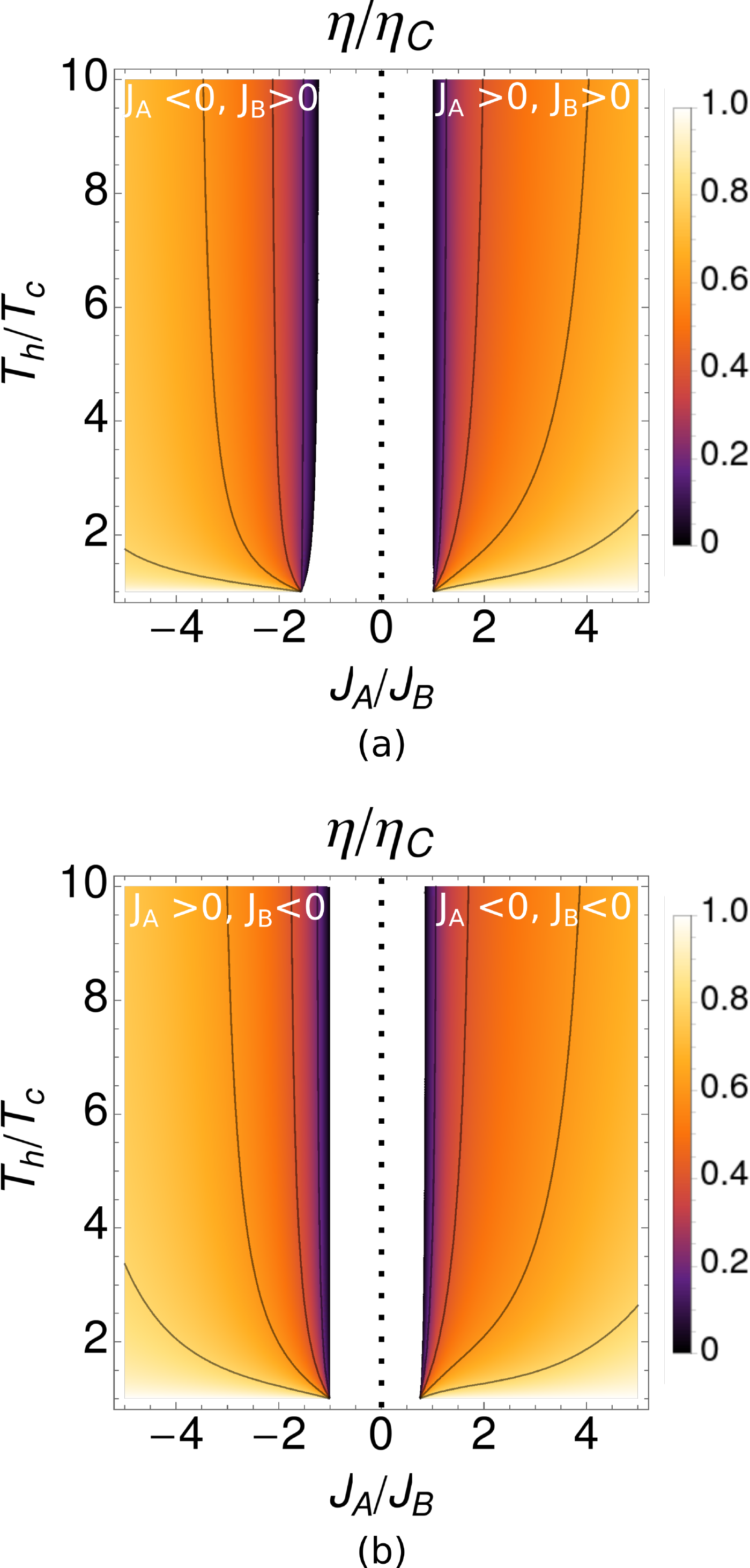}
    \caption{(Color online) Rescaled efficiency $\eta/\eta_{C}$ for the quantum Stirling Heat Engine in terms of the ratio of the parameters of the engine: temperatures ($T_{h}/T_{c}$)  and the magnetic couplings ($J_{A}/J_{B}$). The $J_{A}/J_{B}$ range is extrapolated in order to highlight the heat engine operation region.}
    \label{fig:eff}
\end{figure}

From Eq. \eqref{efficiency}, it is expected that the efficiency goes to zero as the system approaches the transition from a positive to a negative work. In addition, from the equilibrium thermodynamics, the maximal efficiency of a heat engine is bounded by the Carnot efficiency. In order to see this result, Fig. \ref{fig:eff} shows the ratio between the efficiency of the quantum Stirling cycle, Eq. \eqref{efficiency}, and the Carnot efficiency, Eq. \eqref{carnot}, for the regimes of positive work depicted in Fig. \ref{fig:regions}. Here it is worth highlighting that the coupling ratio presented is based on values of applied pressures to change the magnetic coupling of s-1/2 metal complexes (order of $GPa$) \cite{prescimone2010high,cruz2017influence}. 

As observed in Fig. \ref{fig:eff}, the efficiency depends not just on the temperature of the reservoirs but also on the initial and final magnetic coupling strengths. As can be seen, the efficiency increases as we increase the ratio between the initial and final magnetic coupling.

In addition, as anticipated, the transition between $\Wcal>0$ and $\Wcal<0$ is responsible for the drop of the system's efficiency. 
Meanwhile, as the ratio between the temperatures of the heat baths decreases, the efficiency, Eq. \eqref{efficiency}, approaches the  Carnot efficiency, Eq. \eqref{efficiency}, where at $T_{h}\rightarrow T_{c}$ minimizes the Carnot efficiency. As expected, the efficiencies grow as the Carnot point is approached, as depicted in the contour lines.

\section{Experimental approach} 

The experimental implementation of quantum heat engines in metal complexes is an open topic \cite{myers2022quantum}. However, as reported in this paper, in {the} case of dinuclear spin-1/2 metal complexes, the heat exchanged, and the work performed in one complete cycle are expressed via the experimentally observed magnetic susceptibility of the system.
In order to get some advance for the presented model in a practical point of view,  we consider dinuclear hydroxo-bridged Cu(II) compound with formula: [Cu$_2$(OH)$_2$(bpy)$_2$][BF$_4$]2(bpy = 2,2$'$-bipyridine). Reference \onlinecite{prescimone2010high} shows the experimental study of the magnetic susceptibilities {through High-Pressure Magnetometry.} The compound has a central metallic structure that comprises a diamond-like [Cu(II)$_2$(OH)$_2$] unit, composed of two Cu(II) ions bridged by two m-OH- ligands. {For this particular class of compounds, the relationship seen between the Cu–O–Cu bridging angles and the degree of magnetic coupling are given by Eq. \eqref{Jtheta}, which makes this system an experimental realization of the working substance described on Section \ref{substance}.}

{Before to the high-pressure experiment, the magnetic characteristics of the metal complex are determined at ambient pressure after its synthesis. The sample is then inserted in the gasket of a diamond anvil cell, which comprises of two opposing diamond anvils that can exert intense pressure on the sample by tightening the screws of the anvil \cite{prescimone2010high,PhysRevB.99.014417,fu2020sensitive}. After achieving the required pressure, the sample's magnetic characteristics are measured using a SQUID magnetometer \cite{prescimone2010high,PhysRevB.99.014417}. Finally, the magnetic measurement data is then examined to identify the influence of high pressure on the magnetic susceptibility of the metal complex \cite{prescimone2010high}.}

As shown in reference \onlinecite{prescimone2010high}, applying a hydrostatic pressure resulted in a considerable deformation and modification of the structure in this metal complex{, which is reflected in a change in the magnetic coupling between metallic centers} \cite{prescimone2010high}. As pointed out in Section \ref{substance}, this fact is true, especially for bond distances and angles between the metallic center and the bridging ligands \cite{lloveras2021advances,Romanenko_2022}. In this scenario, structural changes manifest in the magnetic behavior of the complexes, as shown by the magnetic susceptibility measurements reported. {Variable temperature magnetic susceptibility measurements were performed in the temperature range $20-350$ K in an applied static magnetic field of 1 kOe at ambient pressure and 0.84 GPa \cite{prescimone2010high}. The applied magnetic field is low enough to maintain the system at the regime of magnetic susceptibility in the system \cite{mario}.}
The susceptibility data were also described by the molar Bleaney-Bowers magnetic susceptibility, Eq. \eqref{eq:09}, where the authors fitted the magnetic coupling constants for each pressure applied.  They found the coupling $J_{A}/k_{B}=-42$ K for an applied pressure of $P_{A}=0.84$ GPa, and  $J_{B}/k_{B}=-32$ K for the ambient pressure ($P_{B}=10^{-4}$ GPa), i.e., during an expansion process the system will increase the magnetic coupling ($J_{B}<J_{A}$). {In order to use the experimental data in the Quantum Stirling Cycle model reported in this paper, the strokes have to be implemented as quasistatically as possible. However, since no time is involved in the quasi-static model, the static magnetic susceptibility measurements can be used to show that the proposed model is consistent with a real scenario. Each experimental point is obtained in the order of seconds; meanwhile, as pointed out in ref. \cite{moreno2018molecular,gaita2019molecular,moreno2021measuring,coronado2020molecular}, the spin-spin relaxation times for such metal complexes are typical of order $1 \sim 100$ $\mu$s. Thus, the two baths are implemented considering the temperature of the magnetometer since the measurement time is far greater than the relaxation of the system itself.}

\begin{figure}[h!]
    \centering
    \includegraphics[width=8.5cm]{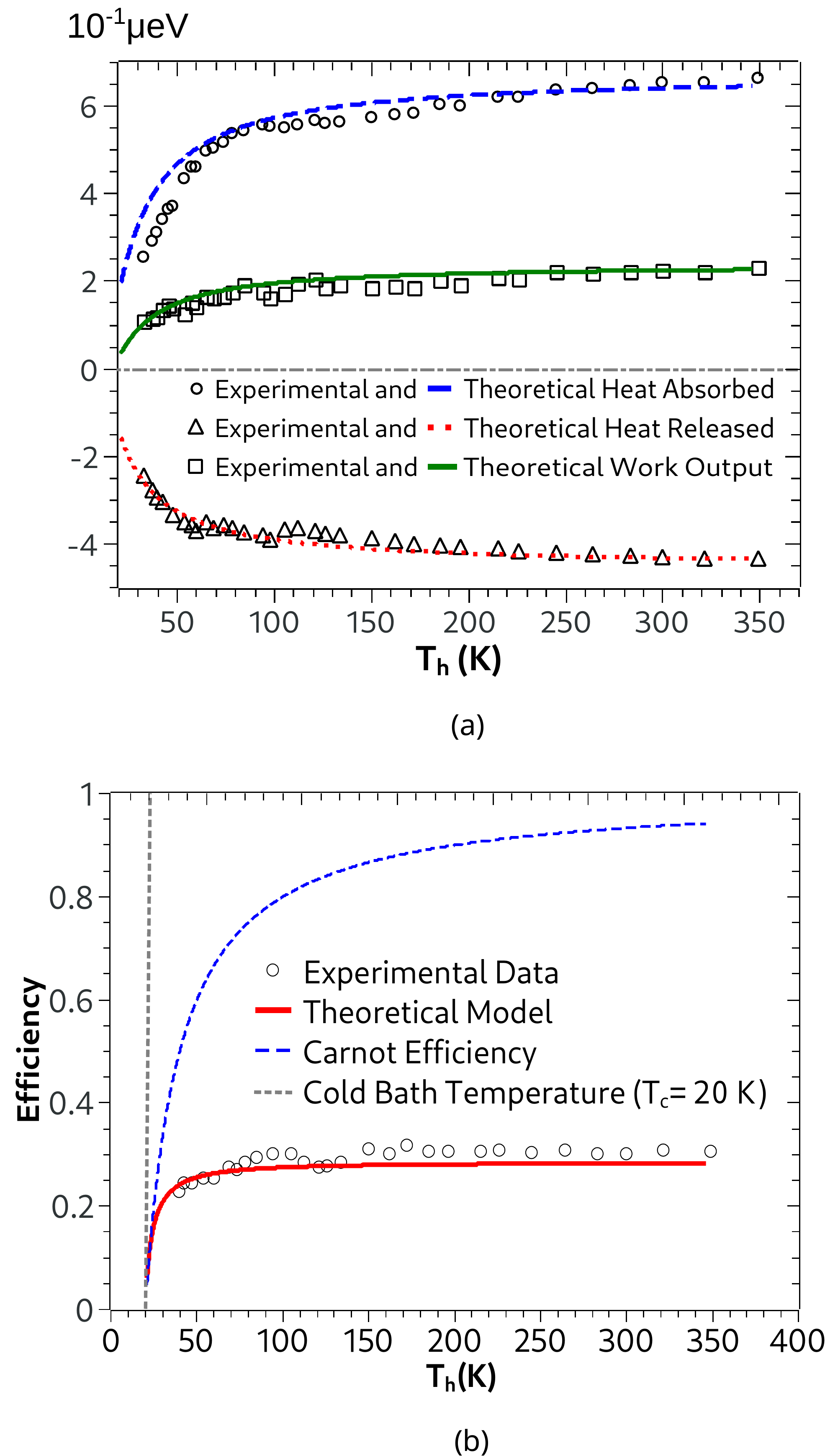}
    \caption{(Color online) {(a) Experimental measurement of work output, Eq. \eqref{work}, heat absorbed, Eq. \eqref{qin}, released, Eq. \eqref{qout} and (b) efficiency, Eq. \eqref{efficiency}, of the quantum Stirling heat engine} as a function of the temperature of the hot reservoir ($T_{h}$), based on the magnetic susceptibility of a dinuclear hydroxo-bridged Cu(II) compound \cite{prescimone2010high}. The engine operates between a hot reservoir at temperature $T_{h}$, and a cold one at fixed temperature  $T_{c} =20$ K (dotted gray line) - the lowest temperature measured for this system \cite{prescimone2010high}. The magnetic couplings are obtained from the Bleaney-Bowers fit of the magnetic susceptibility, where $J_{A}/k_{B}=-42$ K is found for an applied pressure of $P_{A}=0.84$ GPa, and  $J_{B}/k_{B}=-32$ K for the ambient pressure ($P_{B}=10^{-4}$ GPa). The susceptibility-based measurement of the efficiency is in agreement with the theoretical model, shown in Section \ref{operation} (see {Eq}. \ref{efficiency}). As it can be seen, when the temperature of the hot reservoir approaches the cold one, the efficiency of the system approaches the Carnot efficiency (dashed blue line).}
    \label{fig:eff_exp}
\end{figure}

{Applying the Quantum Stirling Cycle model reported in this paper}, one can find, from the operation modes shown in Fig. \ref{fig:regions}, that this system can only operate as a heat engine, regardless of the temperature of the heat baths. Therefore, using the data reported in the reference \onlinecite{prescimone2010high}, one can obtain the dimensionless magnetic susceptibility, Eq. \eqref{eq:09-2}, through the Bleaney-Bowers equation,  Eq. \eqref{eq:09}. Thus, from {Eqs. \eqref{qin} - \eqref{work} and \eqref{efficiency}}, one can obtain the {energy exchanged, as well as the} efficiency of the Quantum Stirling Heat engine from the magnetic susceptibility experimental data. Fig. \ref{fig:eff_exp} shows the experimental measurement of work output, heat absorbed, released, and efficiency in terms of the temperature of the hot reservoir, by fixing the temperature of the cold bath at 20 K - the lowest temperature measured in ref. \onlinecite{prescimone2010high}. 

As it can be seen {in Fig. \ref{fig:eff_exp} (a), the theoretical curves are in agreement with the experimental points obtained from the magnetic susceptibility measurements as shown in ref. \onlinecite{prescimone2010high}. 
This result corroborates the fact that static magnetometry data can be used in a quasi-static scenario, considering the time scale of the measurement and the  spin-spin relaxation times. Furthermore, as anticipated in section \ref{model}, the energy scale of the work output and the heat exchanged obtained from experimental data are in agreement with the results shown in Figs. \ref{fig:qab}-\ref{fig:work_output}, considering the ratios between the magnetic couplings and the temperatures of the thermal baths. In addition, the work output, which is obtained in the order of $10^{-7}$ eV, as shown in Fig. \ref{fig:eff_exp} (a), is several orders of magnitude smaller than the energy scale of the system given by the magnetic couplings (in  the order of $10^{-3}$ eV).}

{Turning to the engine performance, as shown in Fig. \ref{fig:eff_exp} (b),} as the temperature of the hot reservoir approaches the cold one, the measurement of the efficiency approaches the Carnot efficiency (dashed blue line), as expected for the quantum Stirling cycle \cite{chatterjee2021temperature,sur2022,Purkait22} in agreement with the theoretical model (solid red line), also showed in Fig. \ref{fig:eff}. Therefore, the present results are significant in at least two major respects. First, it shows a proof of concept for the theoretical model developed for the quantum Stirling cycle using a dinuclear metal complex as a working substance. In addition, the evaluation of the efficiency of the heat is experimentally accessible through the analytical results shown in this paper. Thus, by measuring a macroscopic property of the system, one can characterize the performance of a quantum heat engine. 

\section{Conclusion}

In summary, this work shows an implementation of a quantum Stirling cycle using the model of a spin-1/2 dinuclear metal complex as a working substance. The system operates between two heat baths with temperatures  $T_{h}$ and $T_{c}$, regarding the hot and cold bath, respectively, and two magnetic couplings  $J_{A}$ and $J_{B}$, induced by the applying of an external hydrostatic pressure. This study has illustrated that four operation modes allowed by the second law of thermodynamics can be tuned by controlling the parameters of the cycle, such as the temperature of the thermal baths and the magnetic couplings of the working substance. 

One of the more significant findings that emerges from this study is that the dimensionless magnetic susceptibility of the working substance could be used to characterize the performance of the Heat engine mode of the quantum Stirling cycle. This fact is experimentally shown through a proof of concept, where the susceptibility-based measurement of the efficiency is obtained from dinuclear hydroxo-bridged Cu(II) as a working substance. Thus, {the data obtained from high-pressure magnetometry experiments can be used to determine the work done, heat exchahed and the efficiency of the heat engine. Therefore,} the presented model can provide insights for future research since it can be used in the application of d$^9$ dinuclear metal complexes as {working substances of} quantum heat engines in terms of their operation modes, from experimental values of magnetic susceptibilities for different values of pressure.

Although this study focuses on the quantum Stirling cycle, these findings may well have a bearing on {solid-state} caloric effects since metal complexes belong to the class of advanced materials explored for caloric applications. Thus, the presented theoretical results lay the groundwork for future research into the application of metal complexes as working substances of quantum heat engines toward the development of emerging quantum technologies based on these advanced materials. 

\begin{acknowledgments}
This study was financed in part by the \textit{Coordena\c{c}\~{a}o de Aperfei\c{c}oamento de Pessoal de N\'{i}vel Superior - Brasil} (CAPES) - finance code 001. M. F. Anka thanks FAPERJ for financial support. T.R. de Oliveira was supported by the Brazilian National Institute for Science and Technology of Quantum Information
(INCT-IQ), and by the Air Force Office of Scientific Research under award number FA9550-19-1-0361. 
\end{acknowledgments}

\end{document}